\documentclass[aps,pop,twocolumn,superscriptaddress,letterpaper]{revtex4}
\usepackage{mathrsfs}
\usepackage{amsmath}
\usepackage{bm}
\usepackage{graphicx}
\usepackage{hyperref}
\usepackage{CJK}

\begin{document}           
\begin{CJK*}{GB}{gbsn}


\title{Generalized Plasma Dispersion Function: One-Solve-All Treatment,
Visualizations,
and Application to Landau Damping}
\author{Hua-sheng XIE (л»ªÉú)}
\email[]{Electronic mail: huashengxie@gmail.com}
\affiliation{Institute for Fusion Theory and Simulation, Zhejiang
University, Hangzhou, 310027, PRC}
\date{\today}

\begin{abstract}
A unified, fast, and effective approach is developed for numerical
calculation of the well-known plasma dispersion function with
extensions from Maxwellian distribution to almost arbitrary
distribution functions, such as the $\delta$, flat top, triangular,
$\kappa$ or Lorentzian, slowing down, and incomplete Maxwellian
distributions. The singularity and analytic continuation problems
are also solved generally. Given that the usual conclusion
$\gamma\propto\partial f_0/\partial v$ is only a rough approximation
when discussing the distribution function effects on Landau damping,
this approach provides a useful tool for rigorous calculations of
the linear wave and instability properties of plasma for general
distribution functions. The results are also verified via a linear
initial value simulation approach. Intuitive visualizations of the
generalized plasma dispersion function are also provided.
\end{abstract}



\maketitle

\section{Introduction}\label{sec:intro}

In a one-dimensional, one-species, non-relativistic electrostatic
plasma system, the Langmuir wave dispersion relation
is\cite{Nicholson1983}
\begin{equation}\label{eq:dres1d}
    D(k,\omega)=1-\frac{\omega_p^2}{k^2}\int_{C}\frac{\partial f_0/\partial
    v}{v-\omega/k}dv=0,
\end{equation}
where $k$ is the wave vector, $\omega=\omega_r+i\gamma$ is the
frequency, $\omega_p=\sqrt{4\pi n_0q^2/m}$ is the plasma frequency
and $C$ is the Landau integral contour.

For Maxwellian distribution $f_0=F_M$, with
\begin{equation}\label{eq:Fmaxell}
    F_M(v)=\frac{1}{\sqrt{\pi}v_t}e^{-\frac{v^2}{v_t^2}},
\end{equation}
the well-known plasma dispersion function (PDF)
\begin{equation}\label{eq:PDF}
    Z_M(\zeta)=\frac{1}{\sqrt{\pi}}\int_{-\infty}^{\infty}\frac{e^{-z^2}}{z-\zeta}dz,
    ~~~\Im(\zeta)>0,
\end{equation}
with analytic continuation to $\Im(\zeta)\leq0$, is defined by Fried
and Conte\cite{Fried1961}, where $\zeta=\omega/(kv_t)$ and
$z=v/v_t$. Hence, (\ref{eq:dres1d}) is rewritten to
\begin{equation}\label{eq:dres1d2}
    D(k,\omega)=1-\frac{1}{(k\lambda_D)^2}\frac{1}{2}Z'_M(\zeta)=0,
\end{equation}
with also
\begin{equation}\label{eq:ZM}
    Z'_M(\zeta)=-2[1+\zeta Z_M(\zeta)],
\end{equation}
where $\lambda_D=\sqrt{T/m}/\omega_p$ and $v_t=\sqrt{2T/m}$. A
$\sqrt{2}$ difference in the normalizations between $v_t$ and $T$
should be noted.

Analytic properties and numerical approaches for the usual PDF
(\ref{eq:PDF}), which is similarly related to complex error
function, Faddeeva function, or Dawson integral, have been
extensively studied since Fried and Conte\cite{Fried1961}. Good
$Z_M(\zeta)$ function numerical schemes for practical application
can also be easily found. However, in studying other distribution
functions, it should be treated separately. The singularity in real
line and analytic continuation to $\Im(\zeta)\leq0$ usually requires
careful treatment, otherwise it would be confusing and would yield
incorrect results.

A family of distributions, i.e., $\kappa$ distributions or
generalized Lorentzian distributions\cite{Valentini2007}
\begin{equation}\label{eq:Fkappa}
    F_{\kappa}=A_{\kappa}\Big[1+\frac{1}{\kappa}\frac{v^2}{v_t^2}\Big]^{-\kappa},
\end{equation}
with the normalization constant
\begin{equation}\label{eq:Akappa}
    A_{\kappa}=\frac{1}{v_t}\frac{\Gamma(\kappa)}{\Gamma(\kappa-1/2)}\frac{1}{\sqrt{\pi\kappa}},
\end{equation}
are very useful for space and astrophysical plasma and have been
studied intensively since Summers and Thorne\cite{Summers1991},
where $\Gamma$ is the Euler gamma function. Recently,
Baalrud\cite{Baalrud2013} also investigated a semi-infinite integral
for Maxwellian distribution, called the incomplete PDF.

Each author uses his or her own technique to treat the Landau
contour. For instance, Baalrud\cite{Baalrud2013} treated the
incomplete PDF via direct numerical integral and continued fraction
expansion, whereas Hellberg and Mace\cite{Hellberg2002} treated the
$\kappa$ distribution using Gauss hypergeometric function.

\begin{figure}
  \includegraphics[width=8cm]{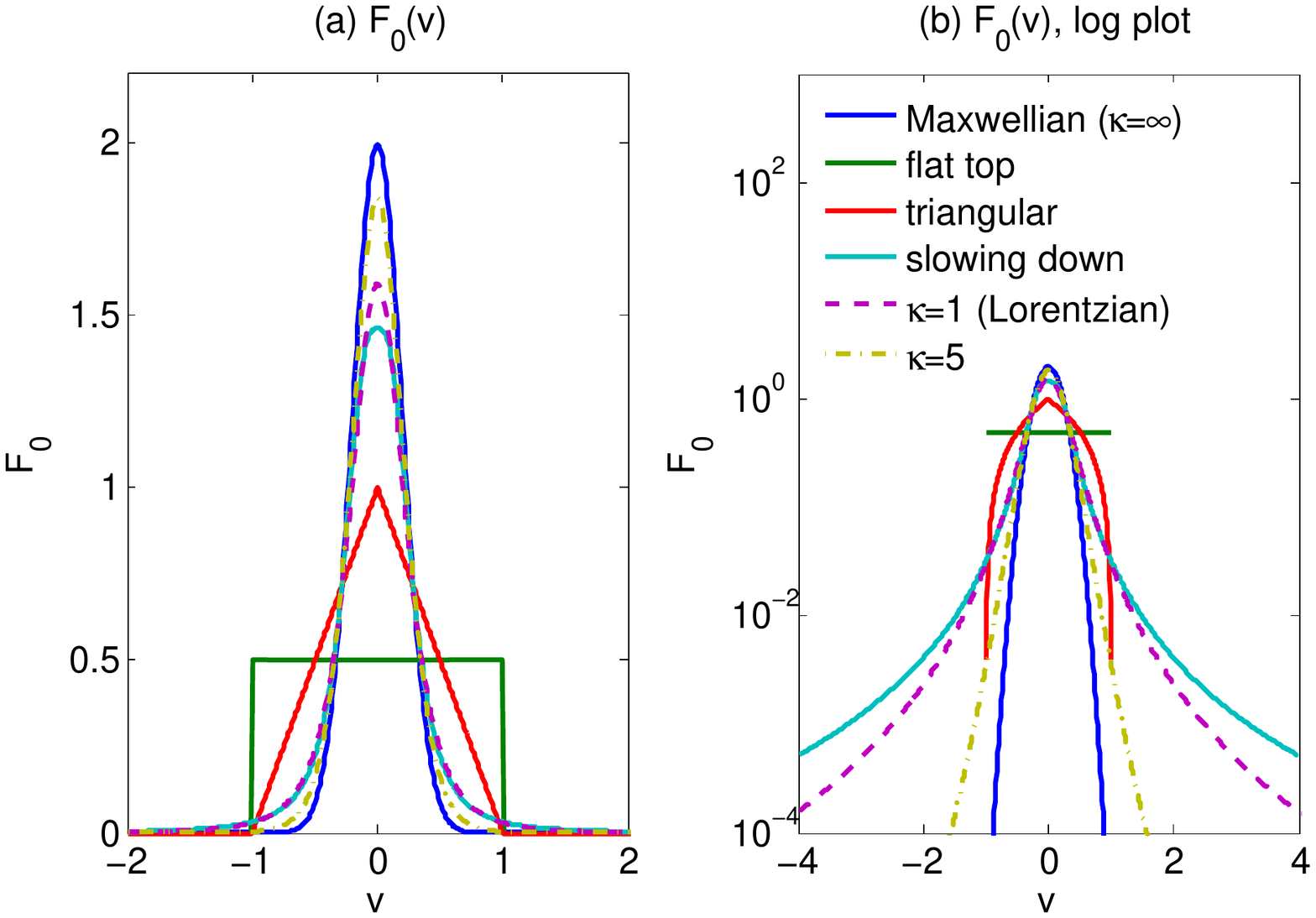}\\
  \caption{Typical distribution functions.}\label{fig:fvexmp}
\end{figure}

The generalized plasma dispersion function (GPDF) can be defined as
\begin{equation}\label{eq:GPDF}
    Z(\zeta)=Z(\zeta,F)=\int_{C}\frac{F}{z-\zeta}dz,
\end{equation}
and its derivative
\begin{equation}\label{eq:GPDFp}
    Z_p(\zeta)=\int_{C}\frac{\partial F/\partial
    z}{z-\zeta}dz=Z'(\zeta,F),
\end{equation}
with the original PDF as a special case when
$F=e^{-z^2}/\sqrt{\pi}$. Developing systematic methods to treat
arbitrary physical reasonable distribution functions (e.g.,
$f\geq0$, $\int fdv<\infty$) in one scheme, i.e., one-solve-all, is
advantageous. For instance, several typical distribution functions
are shown in Fig.\ref{fig:fvexmp}.

In this study, we investigate the analytical properties
(particularly the singularity and analytic continuation problems)
and develop a general numerical scheme for GPDF, i.e., for almost
arbitrary input function $F$.

This problem was also discussed by Robinson\cite{Robinson1990}, who
used the linear combination of orthogonal functions. Three sets of
orthogonal functions, Hermite, Legendre, and Chebyshev polynomials
were discussed. Robinson's method is very similar to our treatment
in this study. However, he has neither given systematic results of
the analytic continuation nor developed a one-solve-all scheme for
practical application.

Our approach, which is discussed in Sec.\ref{sec:meth}, is based on
Hilbert transform (HT) and fast Fourier transform (FFT). After
solving GPDF generally, we show several visualizations of GPDF in
Sec.\ref{sec:visu}. The distribution function effects on Landau
damping are revisited in Sec.\ref{sec:land}, and a summary and
discussion are given in Sec.\ref{sec:summ}.

\section{One-Solve-All Scheme for Generalized Plasma Dispersion Function}\label{sec:meth}

\subsection{Hilbert transform and analytic continuation}
HT is defined as
\begin{equation}\label{eq:Hilb}
    g(z)=H(f(z))=\frac{1}{\pi}\int_{-\infty}^{\infty}\frac{f(z')}{z'-z}dz',
\end{equation}
which can also be viewed as a convolution
\begin{equation}\label{eq:Hilb2}
    g(z)=\frac{1}{\pi z}*f(z),
\end{equation}
or the inverse
\begin{equation}\label{eq:Hilb3}
    f(z)=-\frac{1}{\pi z}*g(z),
\end{equation}
where $f(z)$ and $g(z)$ are called a Hilbert pair. HT usually
represents a $90^\circ$ phase shift of input function.

Some useful properties include
\begin{enumerate}
  \item $H(c_1f_1(z)+c_2f_2(z))=c_1g_1(z)+c_2g_2(z)$,
  \item $H(H(f(z)))=-f(z)$,
  \item $H(f(z+a))=g(z+a)$,
  \item $H(f(az))=sgn(a)g(az)$,
  \item $H(\frac{d^nf(z)}{dz^n})=\frac{d^ng(z)}{dz^n}$,
  \item $f(z)+ig(z)$ is an analytical function.
\end{enumerate}

A good scheme for numerically calculating HT in real line is
provided by Weideman\cite{Weideman1995}. Two other numerical methods
are, using (\ref{eq:Hilb}), i.e.,
\begin{equation}\label{eq:HilbN1}
    g(z)\simeq\frac{2}{\pi}\sum_{n=-\infty}^{\infty}\frac{f(z+(2n+1)h)}{2n+1},
\end{equation}
where $h$ is the step size, or using (\ref{eq:Hilb2}), i.e.,
\begin{equation}\label{eq:HilbN2}
    g(z)=ift[ft(\frac{1}{\pi z})\cdot ft(f(z))],
\end{equation}
where $ft()$ and $ift()$ denote the Fourier transform and its
inverse.

The methods in Weideman's 1995 paper\cite{Weideman1995} or via
(\ref{eq:HilbN1}) and (\ref{eq:HilbN2}) are mainly for calculating
integral principal value (PV) in real line.

In fact, the definition of (\ref{eq:Hilb}) is not a single function.
For simplification, we require $f(z)$ be an entire function that is
integrable in the range of $-\infty$ to $+\infty$. The plasma
dispersion function is defined for $\Im(z)>0$, and thus, should be
extended to $\Im(z)\leq0$, which is\cite{Jones1985}
\begin{equation}\label{eq:HilbC}
    g^{+}(z)=\left\{
    \begin{array}{ccc}
    \frac{1}{\pi}\int_{-\infty}^{\infty}\frac{f(z')}{z'-z}dz',~&\Im(z)>0,\\
    \frac{1}{\pi}PV\int_{-\infty}^{\infty}\frac{f(z')}{z'-z}dz'+if(z),~&\Im(z)=0,\\
    \frac{1}{\pi}\int_{-\infty}^{\infty}\frac{f(z')}{z'-z}dz'+2if(z),~&\Im(z)<0.
    \end{array}\right.
\end{equation}

For instance, the HT of Lorentzian distribution
$f(z)=\frac{a}{\pi}\frac{1}{z^2+a^2}$ is
\begin{equation}\label{eq:LorentzHT}
    g(z)=\left\{
    \begin{array}{ccc}
    -\frac{1}{\pi}\frac{1}{(z+ia)},~&\Im(z)>0,\\
    -\frac{z}{(z^2+a^2)},~&\Im(z)=0,\\
    -\frac{1}{\pi}\frac{1}{(z-ia)},~&\Im(z)<0,
    \end{array}\right.
\end{equation}
whereas
\begin{equation}\label{eq:ZgLorentz}
  g^{+}(z)=-\frac{1}{\pi}\frac{1}{(z+ia)},
\end{equation}
which is consistent with (\ref{eq:HilbC}).

$g^{-}(z)$ can be defined in a similar manner if we want to extend a
function from lower half plane to the entire complex plane.

Weideman\cite{Weideman1994} also provided a method to calculate
$g^{+}(z)$ of the HT of Gaussian function in upper half plane, which
is related to PDF $Z_M(\zeta)$.

\subsection{One-solve-all approach}
A comparison of (\ref{eq:GPDF}) and (\ref{eq:GPDFp}) with
(\ref{eq:HilbC}) indicates $Z=\pi g^{+}$ and $Z_p=\pi g'^{+}$ with
$F=f$ and $F'=f'$. Thus, GPDF is merely a HT of distribution
function and shares the same properties of HT as listed in the above
subsection.

Jones {\it et al.}\cite{Jones1985} also discussed the contour
integral problem of GPDF and used transformation $z=tan(t)$ to map
the integral of $z\in(-\infty,\infty)$ to $t\in(-\pi,\pi)$. This
method is merely an alteration of (\ref{eq:HilbN1}) and requires
other tricks to avoid singularity. Moreover, the method is time
consuming and not suitable for high accuracy calculation as the
discrete step should be very small to avoid divergence. Another
direct numerical integral result is shown by Guio {\it et
al.}\cite{Guio1998} with typical errors of $10^{-4}$.

In this paper, we use (\ref{eq:HilbC}) to extend Weideman's
approach\cite{Weideman1994,Weideman1995} to the entire complex
plane, where the orthogonal functions is $e^{i\theta}$, which can be
evaluated very rapidly by FFT, instead of the Hermite, Legendre, and
Chebyshev polynomials used by Robinson\cite{Robinson1990}.

The key steps are summarized as follows.

Assuming an expansion
\begin{equation}\label{eq:expan}
    [W(v)]^{-1}F(v)=\sum_{n=-\infty}^{\infty}a_n\rho_n(v),~~~~v\in\mathcal{R}
\end{equation}
where $\{\rho_n(v)\}$ is an orthogonal basis set with weight
function $W(v)$, i.e.,
\begin{equation}\label{eq:basis0}
    \int_{-\infty}^{\infty}W(v)\rho_n(v)\rho_m^*(v)dv=A\delta_{m,n},
\end{equation}
where the asterisk denotes complex conjugation and $\delta_{m,n}$ is
the Kronechker delta. The coefficients are
\begin{equation}\label{eq:expanan}
    a_n=\frac{1}{A}\int_{-\infty}^{\infty}F(v)\rho_n^*(v)dv.
\end{equation}
Then
\begin{equation}\label{eq:expan2}
    \frac{F(v)}{v-z}=\sum_{n=-\infty}^{\infty}a_n\Big[W(v)\frac{\rho_n(v)}{v-z}\Big].
\end{equation}

For the upper half plane, we use weight function
$W(v)=1/(L^2+v^2)$\cite{Weideman1994} and basis functions
\begin{equation}\label{eq:basis}
    \rho_n(v)=\frac{(L+iv)^n}{(L-iv)^{n}},
\end{equation}
which is a Fourier form because $e^{i\theta}=(L+iv)/(L-iv)$ with
$v=L\tan(\theta/2)$ and $dv/d\theta=(L^2+v^2)/(2L)$, then $a_n$ can
be evaluated using FFT and we can obtain $A=\pi/L$ using
$\int_{-\pi}^{\pi}e^{in\theta}e^{-im\theta}d\theta=2\pi\delta_{m,n}$.

Using residues, for $\Im(\zeta)>0$, we find the integrals
\begin{equation}\label{eq:intgup}
    \int_{-\infty}^{\infty}\frac{W(v)}{v-z}\frac{(L+iv)^n}{(L-iv)^n}dv=\left\{
    \begin{array}{ccc}
    \frac{i\pi}{L}\frac{1}{(L-iz)},~&n=0,\\
    \frac{2i\pi}{L^2+z^2}\frac{(L+iz)^n}{(L-iz)^n},~&n>0,\\
    0,~&n<0.
    \end{array}\right.
\end{equation}
We obtain
\begin{equation}\label{eq:gup}
    g^+(z)=\frac{2i}{L^2+z^2}\sum_{n=1}^{\infty}a_n\Big(\frac{L+iz}{L-iz}\Big)^n
    +\frac{ia_0}{L(L-iz)},~~~\Im(z)>0.
\end{equation}

For the lower half plane, $\Im(\zeta)<0$, we use $F(z)=[F(z^*)]^*$
and obtain
\begin{equation}\label{eq:glow}
    Z(\zeta)=[Z(\zeta^*)]^*+2i\pi f(\zeta),~~~\Im(\zeta)<0.
\end{equation}

For real line, $\Im(\zeta)=0$, we use $W(v)=1$ and
$\rho_n(v)=(L+iv)^n/(L-iv)^{n+1}$\cite{Weideman1995}, and obtain
\begin{subequations}\label{eq:greal}
\begin{eqnarray}
    g^+(z) &=& \sum_{n=-\infty}^{\infty}isgn(n)a_n\rho_n(z)+if(z)\\
           &=& \sum_{n=0}^{\infty}2ia_n\rho_n(z),~~~\Im(z)=0.
\end{eqnarray}
\end{subequations}

We have avoided the singularity on real line by treating the
integrals of the basis functions analytically.

A completed scheme\footnote{See supplementary material at [URL will
be inserted by AIP] for the MATLAB version numerical routines, where
GPDF, root finding, IVP simulation and test plotting codes are
included.} is provided through the combination of (\ref{eq:gup}),
(\ref{eq:glow}) and (\ref{eq:greal}), which can support an almost
arbitrary smooth distribution function $F(v)$ and $F_p(v)=\partial
F/\partial v$ as input function.

In numerical calculation, we truncate the summation at a finite
point $n=N$. In the practical test for Gaussian input function with
$N=32$, the program delivers twelve significant decimal
digits\cite{Weideman1994}, where $L$ is an optimal parameter and is
set to $2^{-1/4}N^{1/2}$ as default.

Another good feature of this approach is that the coefficients $a_n$
need only be calculated once for all $z$, making the scheme even
faster.

Moreover, an unexpected but interesting feature is that, the input
function in the L.H.S. of (\ref{eq:expan}) is not necessarily a
smooth function. The R.H.S. of (\ref{eq:expan}) can transform the
real function $F(v)$ in real line to a smooth analytic complex
function in whole complex plane with truncation. This feature can
help us calculate several (note: not all) non-smooth input functions
directly, such as flat top, incomplete Maxwellian, and slowing down.
The validity of this approximation will be verified in
Sec.\ref{sec:visu}.

\section{Visualizations of Generalized Plasma Dispersion Function}\label{sec:visu}

\subsection{Maxwellian distribution}
First, we compare the results of the usual PDF with Maxwellian
distribution as input function.

\begin{figure}
  \includegraphics[width=8cm]{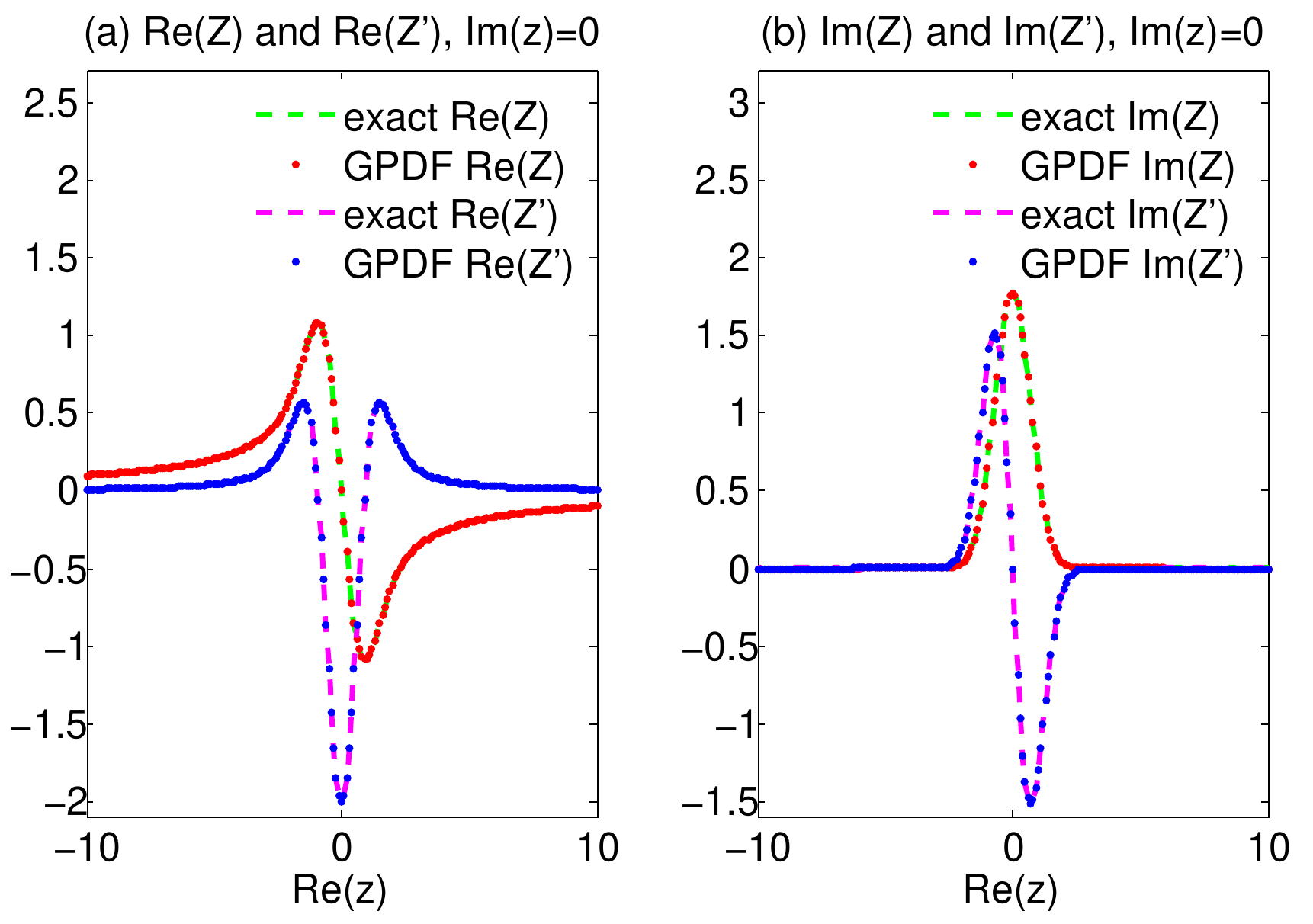}\\
  \caption{Comparison of our scheme for GPDF with exact $Z(\zeta)$ function via
  complex error function in standard numerical library on real axis.}\label{fig:ZMaxwell1D}
\end{figure}

\begin{figure}
  \includegraphics[width=8cm]{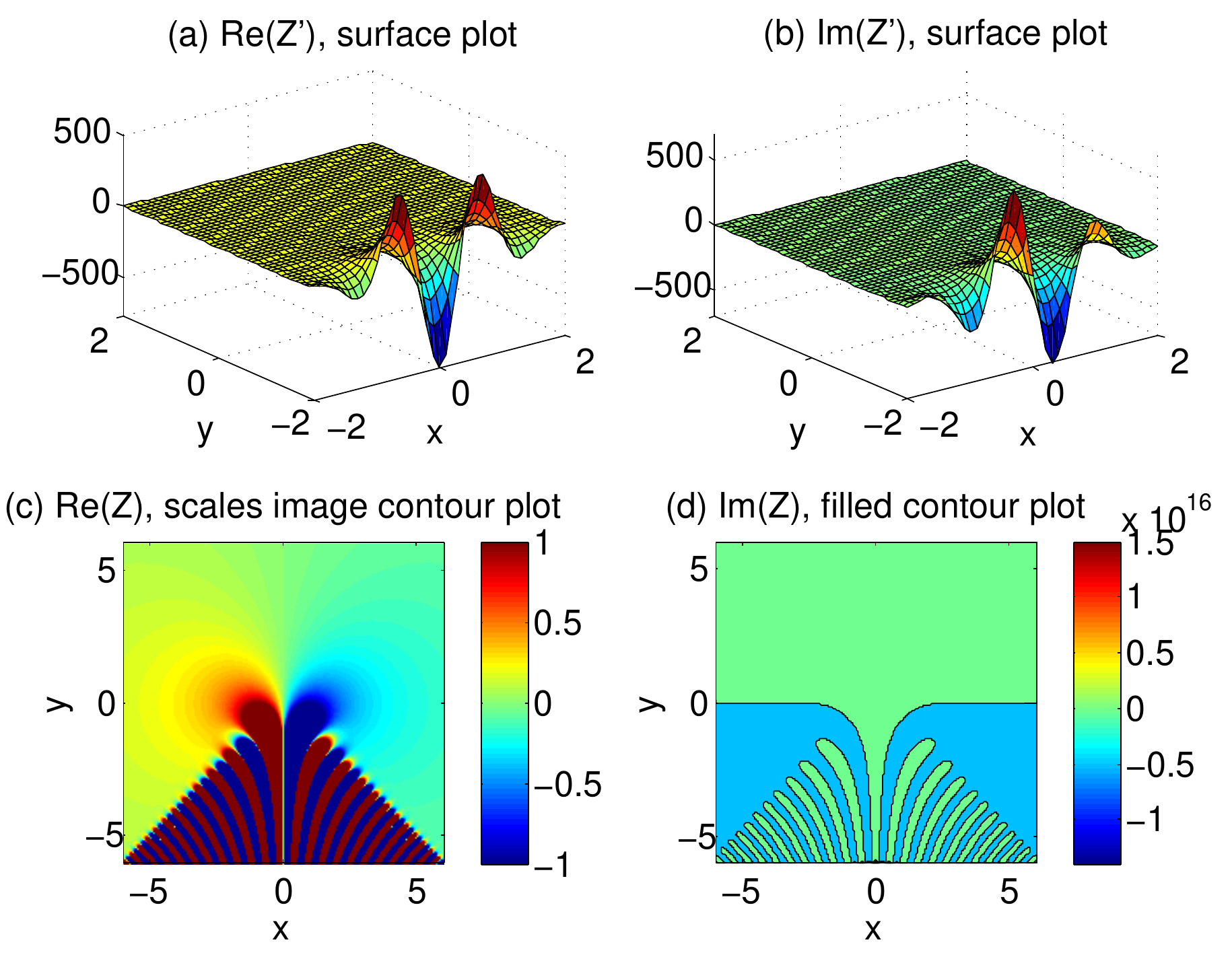}\\
  \caption{Visualization of $Z(\zeta)$ and $Z'(\zeta)$ with Maxwellian input function.}\label{fig:ZMaxwell}
\end{figure}

Fig.\ref{fig:ZMaxwell1D} shows a comparison of our scheme (using
$N=32$) with exact $Z(\zeta)$ function via complex error function in
standard numerical library on real axis. We find the errors to be
around $10^{-12}$ (not shown in the figure), e.g.,
$Z(1)=-1.076159013825734 + 0.652049332173291i$ in our scheme and
$Z(1)=-1.076159013825537 + 0.652049332173292i$ via standard library.

Fig.\ref{fig:ZMaxwell} shows the 2D visualizations of $Z(\zeta)$ and
$Z'(\zeta)$ produced by our scheme, which shows that the functions
are indeed analytically smooth. If we exclude the step for analytic
continuation, we will find a jump at real line $\Im(z)=0$ (not shown
here).

\subsection{$\kappa$ distribution}
\begin{figure}
  \includegraphics[width=8cm]{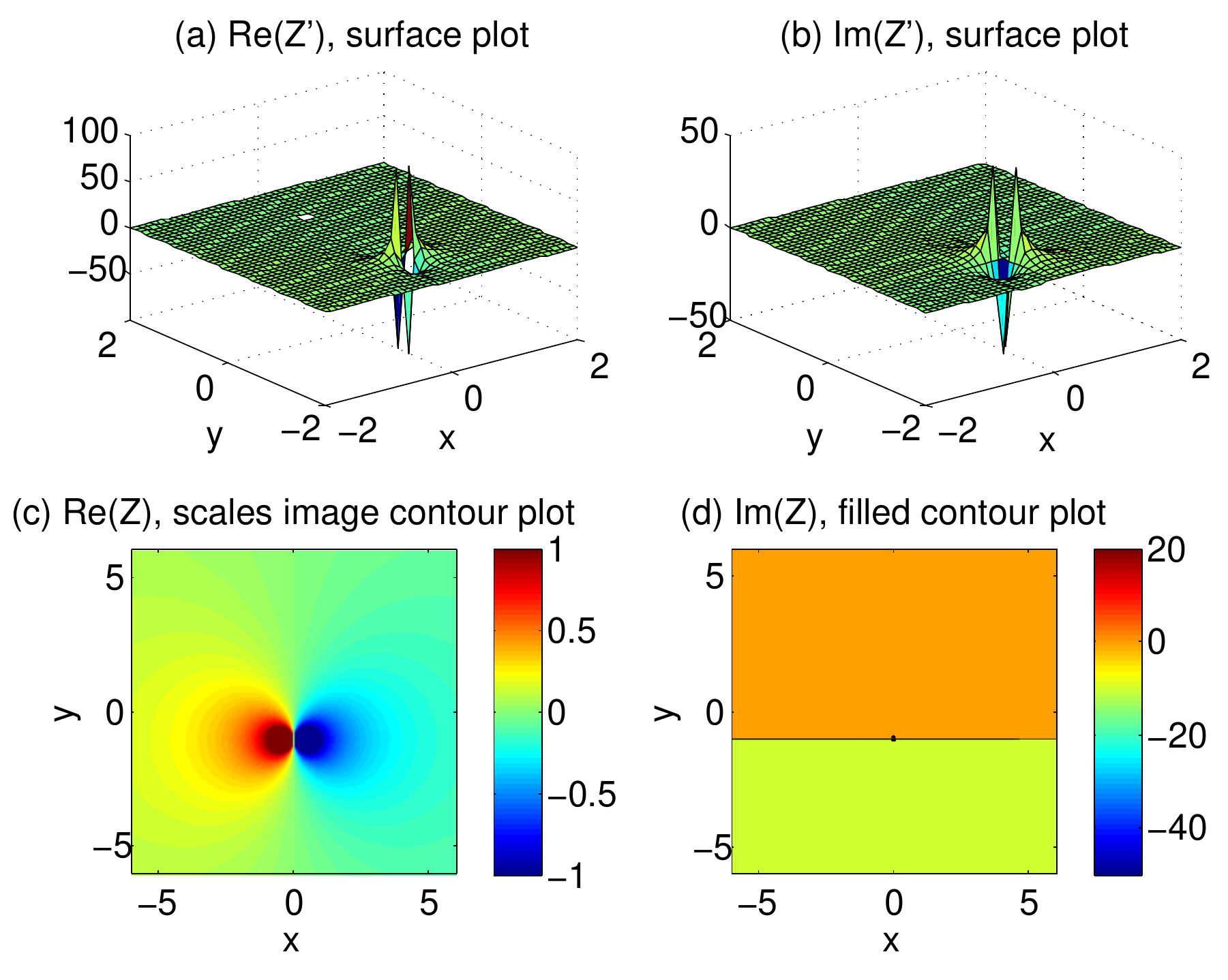}\\
  \caption{Visualization of $Z(\zeta)$ and $Z'(\zeta)$ with input function
  $F=\frac{1}{\pi(v^2+1)}$, where artificial singular point {\it NaN} at
  $\zeta=i$ is kept.}\label{fig:ZLorentz}
\end{figure}

With $F=\frac{1}{\pi(v^2+1)}$ as input function, a result is shown
in Fig.\ref{fig:ZLorentz}, where we can find a singularity at
$z=-i$, which is consistent with analytical result
(\ref{eq:ZgLorentz}).

However, another artificial singular point at $z=i$ [see panel (a)]
can be found, where the code yields {\it NaN}. The two first-order
singular points $v-ia=0$ and $v-z=0$ transformed into one
second-order singular point $(v-ia)^2=0$ when $z=ia$ for Lorentzian
input function. An extra approximation $Z(\zeta_0)\simeq
Z(\zeta_0+\epsilon)$ in the code is used to treat this kind of
singular point, with $\zeta_0$ as the artificial singular point and
$\epsilon$ be a small number, e.g., $10^{-10}$. After the extra
treatment, we find GPDF yields exactly the same values as
$\pi\times$(\ref{eq:ZgLorentz}) in all computation grids (not shown
here) with controllable small errors.

\begin{figure}
  \includegraphics[width=8cm]{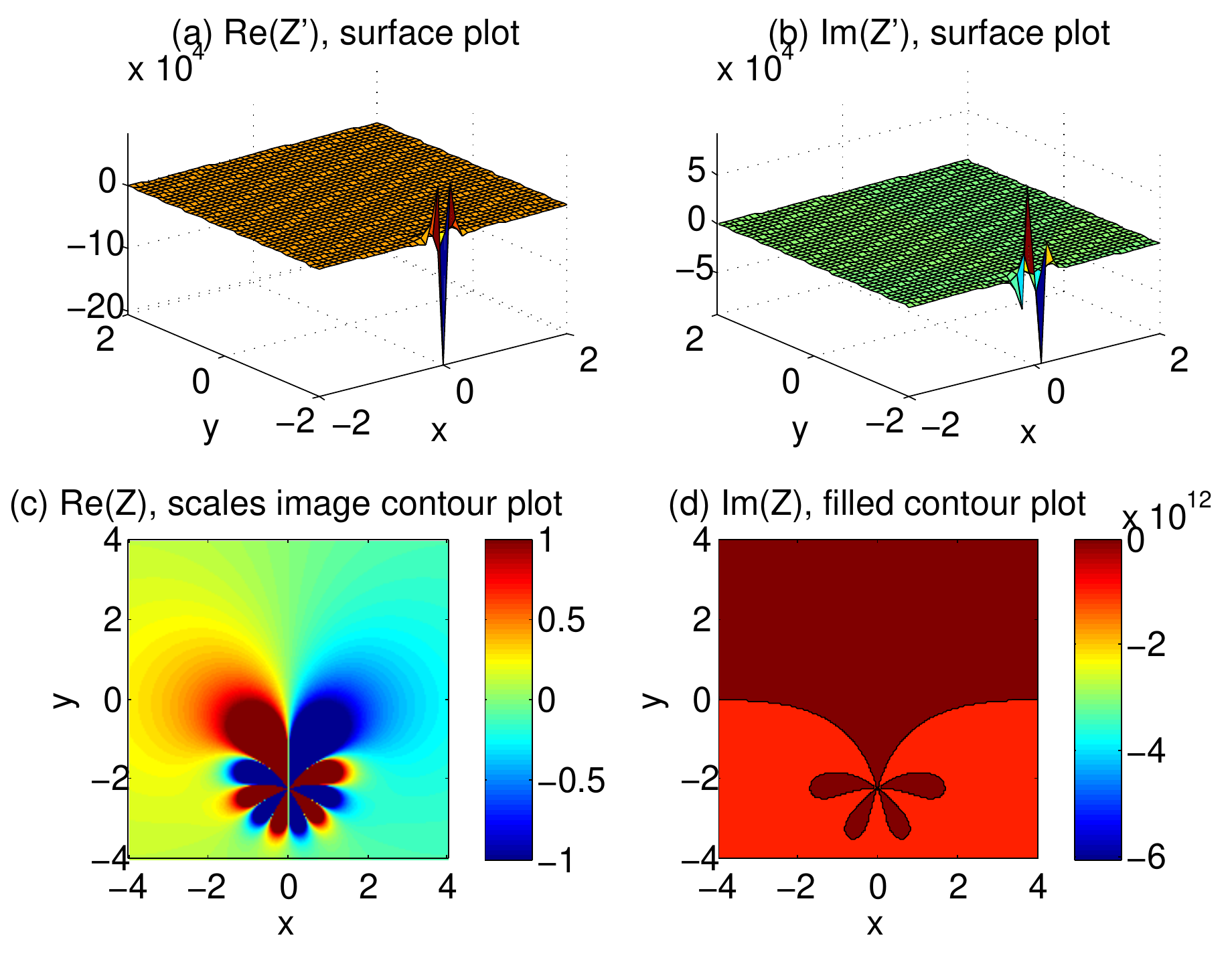}\\
  \caption{Visualization of $Z(\zeta)$ and $Z'(\zeta)$ with input function $F_{\kappa=5}$.}\label{fig:ZKappa5}
\end{figure}

After fixing this problem, a result is shown in
Fig.\ref{fig:ZKappa5} for $\kappa=5$, $v_t=1$.

To keep $\kappa=1$ the usual Lorentzian distribution, the definition
of $\kappa$-distribution in (\ref{eq:Fkappa}) slightly differs from
the usual one\cite{Summers1991} but is close to the one by Valentini
and D'Agosta\cite{Valentini2007}.

\subsection{$\delta$ distributions}

The scheme described in Sec.\ref{sec:meth} cannot support several
non-standard distribution functions directly, particularly, $\delta$
distribution, which has been widely used for modeling cold plasma.
We treat it separately in the code, via analytical expression.

\begin{figure}
  \includegraphics[width=8cm]{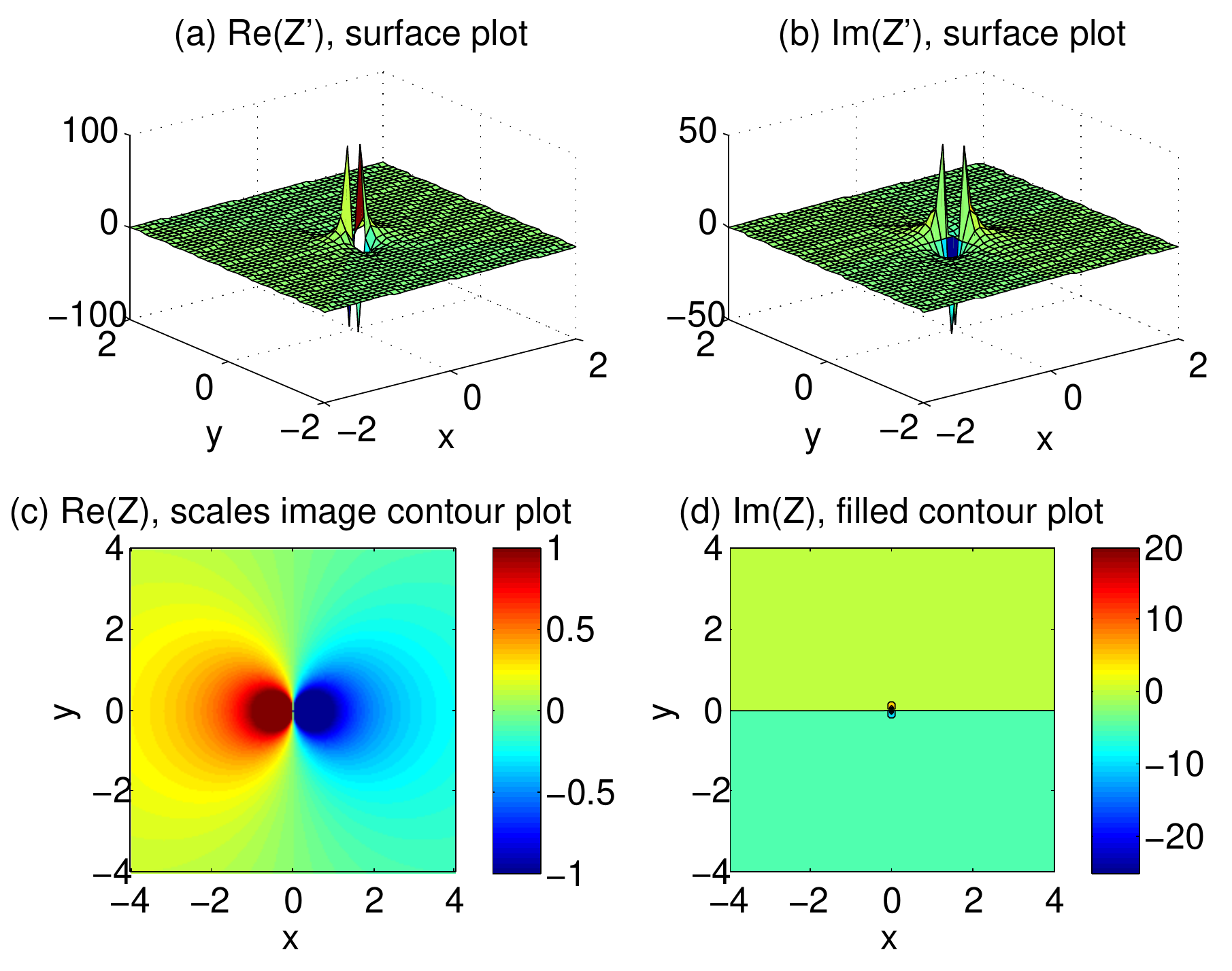}\\
  \caption{Visualizations of $Z(\zeta)$ with input function $F_{\delta}$
  and parameters $z_d=0$.}\label{fig:ZDelta}
\end{figure}

For $\delta$-distribution $F_{\delta}=\delta(z-z_d)$,
\begin{subequations}\label{eq:Zdelta}
\begin{eqnarray}
    Z(\zeta) &=& -\frac{1}{\zeta-z_d},\\
    Z_p(\zeta) &=& \frac{1}{(\zeta-z_d)^2},
\end{eqnarray}
\end{subequations}
which can also be obtained from (\ref{eq:ZgLorentz}) using the limit
$a\to0$ because
$\delta(v)=\lim\limits_{a\to0}\frac{1}{\pi}\frac{a}{v^2+a^2}$. The
result is shown in Fig.\ref{fig:ZDelta}.

\subsection{Incomplete Maxwellian distributions}
The input function is
\begin{equation}\label{eq:FIMaxell}
    F_{\text{IM}}(v)=H(v-\nu)\frac{1}{\sqrt{\pi}}e^{-v^2},
\end{equation}
which has been investigated comprehensively by
Baalrud\cite{Baalrud2013}, where $H$ is the Heaviside step function.

As previously mentioned, the numerical scheme cannot treat the
$\delta$ function directly. For $Z'$, an extra correction term
should be added, i.e.,
$Z'_{\text{IM}}(\zeta)=Z'_{\text{IM0}}(\zeta)-\frac{1}{\sqrt{\pi}}e^{-\nu^2}/(\zeta-\nu)$,
where $Z'_{\text{IM0}}$ is the result without correction.

\begin{figure}
  \includegraphics[width=8cm]{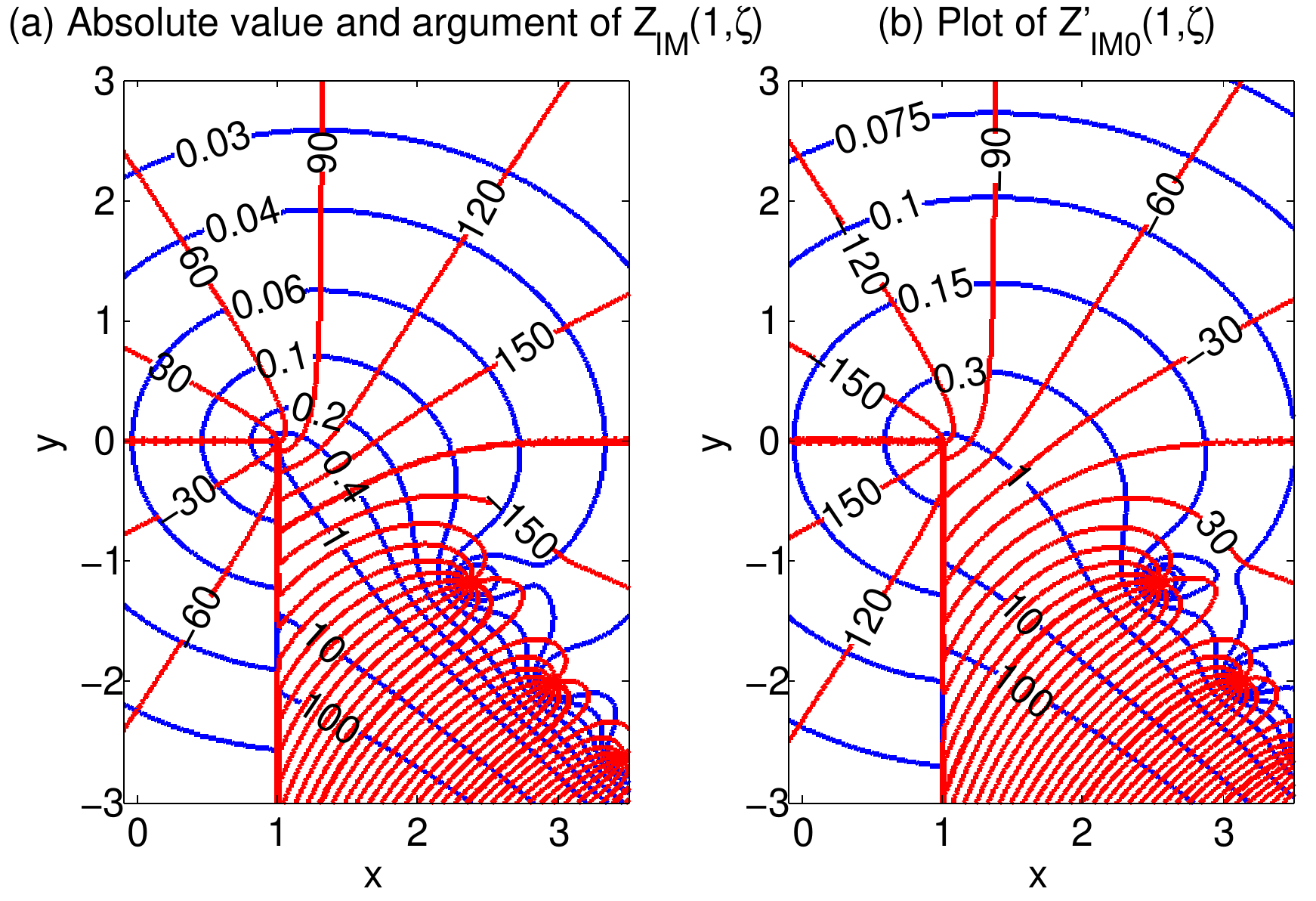}\\
  \caption{Visualization of $Z_{\text{IM}}(\zeta)$ and $Z'_{\text{IM0}}(\zeta)$ with
  input function (\ref{eq:FIMaxell}), where the absolute value and argument are calculated using $Z=|Z|e^{i\theta}$.}\label{fig:ZIMA}
\end{figure}

Fig.\ref{fig:ZIMA} shows the results of $Z_{\text{IM}}(\zeta)$ and
$Z'_{\text{IM0}}(\zeta)$ calculated from our GPDF scheme, with
$\nu=1$, $N=1024$ and $L=10$. Fig.\ref{fig:ZIMA}(a) is in favor of
Fig.3(b) by Baalrud\cite{Baalrud2013}, which was calculated from a
direct numerical integral. Our one-solve-all scheme can solve the
same problem at least ten times faster than the direct numerical
integral usually with the same accuracy.

A slight error can be found around the real line when $N$ is small,
say $N=64$, which arises from the error of Fourier expansion around
the sharp step place (Gibbs phenomenon) that requires a large $N$ to
overcome.

\subsection{Flat top distribution}

For flat top distribution
$F_{\text{Rect}}=\frac{H(z-z_a)-H(z-z_b)}{z_b-z_a}$,
\begin{subequations}\label{eq:Zrect}
\begin{eqnarray}
    Z(\zeta) &=& \frac{1}{z_b-z_a}\Big[\ln(\zeta-z_b)-\ln(\zeta-z_a)\Big],\\
    Z_p(\zeta) &=&
    \frac{1}{z_b-z_a}\Big[\frac{1}{\zeta-z_b}-\frac{1}{\zeta-z_a}\Big].
\end{eqnarray}
\end{subequations}

\begin{figure}
  \includegraphics[width=8cm]{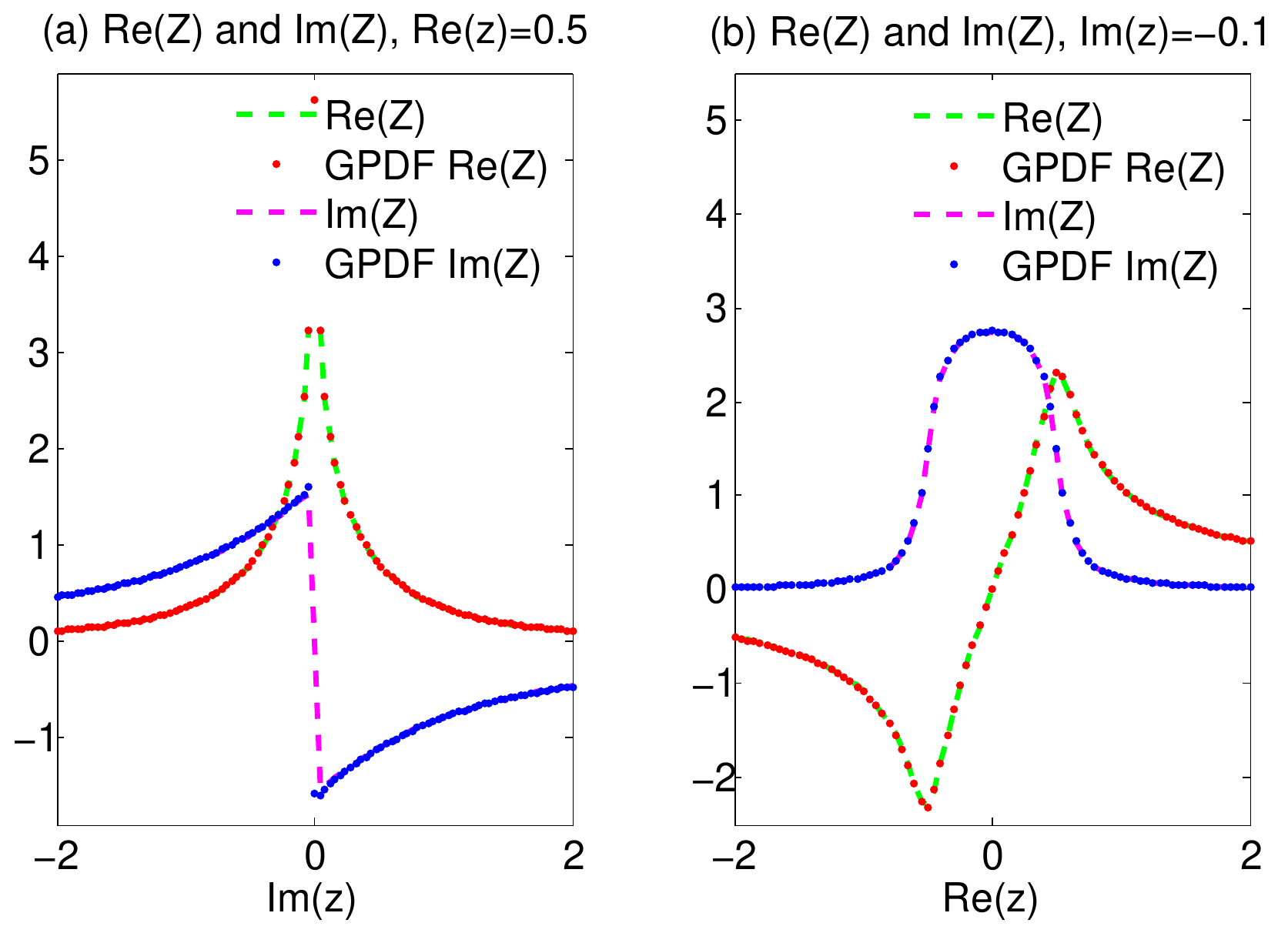}\\
  \caption{Comparison of GPDF results and analytical results for $Z(\zeta)$ with input function
  $F_{\text{Rect}}$ and parameters $z_a=-0.5$ and $z_b=0.5$.}\label{fig:ZRect}
\end{figure}

GPDF results and analytical results are compared in
Fig.\ref{fig:ZRect}. This benchmark indicates that our treatment for
GPDF is indeed suitable for both smooth and non-smooth input
function. However, to keep the result exactly the same as
(\ref{eq:Zrect}) at range $z_a \leq \zeta \leq z_b$ for lower half
plane [$\Im(\zeta)<0$], the analytic continuation term $2if(z)$ is
set to zero, instead of $2i/(z_b-z_a)$.

For $Z'_{\text{Rect}}$, we need also an extra treatment because of
the $\delta$ function.

\subsection{Triangular distribution}
\begin{figure}
  \includegraphics[width=8cm]{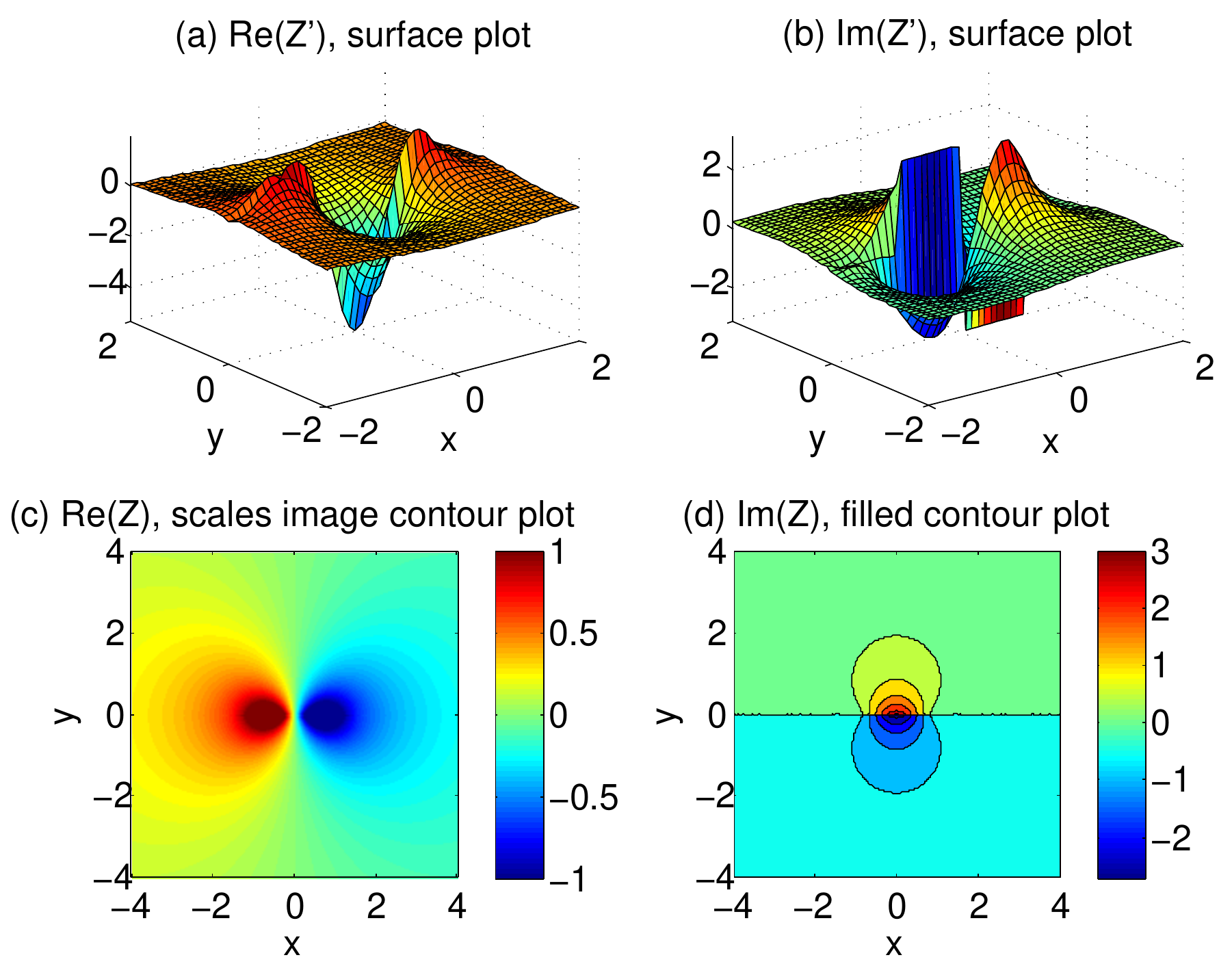}\\
  \caption{Visualization of $Z(\zeta)$ and $Z'(\zeta)$ with input function $F_{\text{Tri}}$
  and parameters $z_a=-1.0$, $z_b=0$ and $z_c=1$.}\label{fig:ZTri}
\end{figure}

The distribution function is
\begin{equation}
\begin{split}\label{eq:FTri}
    F_{\text{Tri}}&=\frac{H(z-z_a)-H(z-z_b)}{(z_b-z_a)}\frac{2}{z_c-z_a}(z-z_a) \\
           & -\frac{H(z-z_b)-H(z-z_c)}{(z_c-z_b)}\frac{2}{z_c-z_a}(z-z_c),
\end{split}
\end{equation}
and corresponding $Z(\zeta)$ is shown in Fig.\ref{fig:ZTri}.

The analytic continuation term $2if(z)$ for lower half plane
[$\Im(\zeta)<0$] is set to zero. Comparing root finding results and
simulation results of Langmuir wave using the methods discussed in
Sec.\ref{sec:land}, we find that the solutions are the same (not
shown here) for both cases, i.e., setting the term to zero or
non-zero, and agreed with the simulations.

\subsection{Slowing down distributions}
The distribution function is very common in fusion plasma, such as
tokamak, for fast particles,
\begin{equation}\label{eq:FSD}
    F_{\text{SD}}=\frac{3\sqrt{3}v_t^2}{4\pi}\frac{1}{|v|^3+v_t^3}H(v_c-|v|).
\end{equation}

\begin{figure}
  \includegraphics[width=8cm]{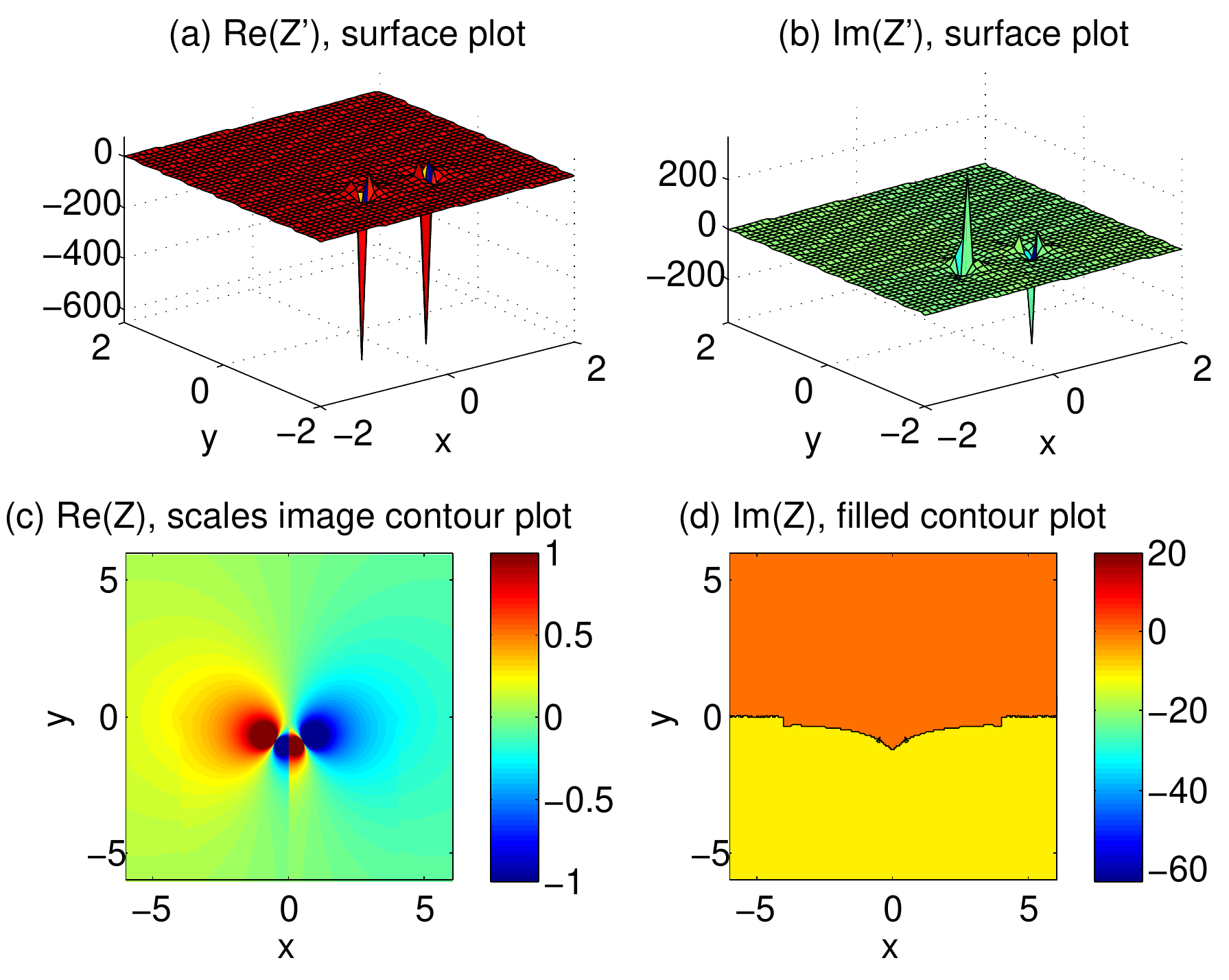}\\
  \caption{Visualization of $Z(\zeta)$ and $Z'(\zeta)$ with input function $F_{\text{SD}}$ for $v_t=1$ and $v_c=4$.}\label{fig:ZSD}
\end{figure}

A result is shown in Fig.\ref{fig:ZSD}.

One should note that the absolute value of $v$ in $F$ could cause
problems in the complex plane, as
$|\Re(v)+i\Im(v)|=\sqrt{\Re(v)^2+\Im(v)^2}\neq|\Re(v)|+i\Im(v)$. We
use $H(v)$ to rewrite $|v|$ in the code, where
$H(\Re(v)+i\Im(v))=H(\Re(v))[\Re(v)+i\Im(v)]$. Hence, we can still
use $if(z)$ directly for the analytic continuation term to reduce
numerical errors, instead of using the expansion expression $\sum
a_n\rho_n$.

\subsection{Other distributions}

\begin{figure}
  \includegraphics[width=8cm]{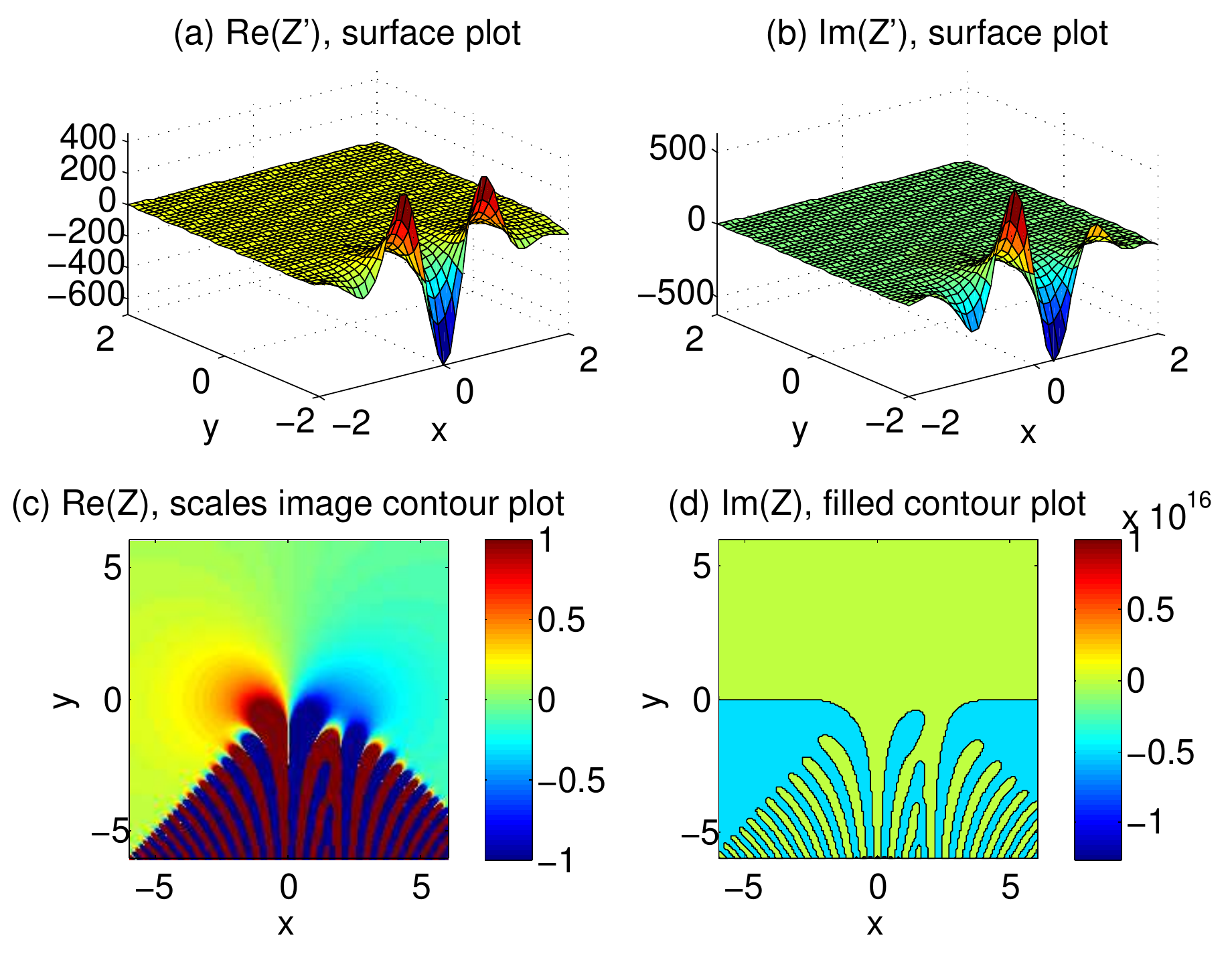}\\
  \caption{Visualization of $Z(\zeta)$ and $Z'(\zeta)$ with bump-on-tail input function.}\label{fig:ZBeam}
\end{figure}

The GPDF has several good features such as for the bump-on-tail
problem, when using usual PDF, the core plasma and beam plasma
should to be treated separately. We can treat it using one input
function directly when using GPDF. A result is shown in
Fig.\ref{fig:ZBeam} for
$F=0.9e^{-v^2}/\sqrt{\pi}+0.1e^{-(v-2)^2}/\sqrt{\pi}$. A comparison
with Fig.\ref{fig:ZMaxwell} shows the beam tail affects $Z(\zeta)$
apparently, particularly at the place around $\Re(\zeta)=v_d=2$.

\subsection{Short summary}

The one-solve-all scheme has been shown to be effective. However, to
treat non-smooth/analytical input functions, extra corrections
should be noted. For non-smooth flat top and triangular
distributions, the method of determining the analytic continuation
requires further investigation, because it cannot be distinguished
by Langmuir wave simulation.

From the visualizations of $Z(\zeta)$ and $Z'(\zeta)$ for different
types of input functions, the quantitative value, shape, or topology
of $Z$ and $Z'$ vary considerably, which will then bring different
kinetic effects, e.g., Landau damping rate.

\section{Distribution Function Effects on Landau Damping}\label{sec:land}

\subsection{Benchmark GPDF using initial value scheme}
For the initial value scheme, the starting equations are the
normalized linear Vlasov-Poisson equations
\begin{equation} \label{eq:es1d}
\left\{\begin{aligned}
{\partial_t\delta f}&=-ikv\delta f+\delta E{\partial_v{f_0}}, \\
ik\delta E &=-\int{\delta fdv}.
\end{aligned} \right.
\end{equation}
We usually set $\lambda_D=1$, then $v_t=\sqrt{2}$ in the initial
distribution function $f_0$, e.g., $f_0=\exp(-v^2/2)/\sqrt{2\pi}$
for Maxwellian.

Eqs.(\ref{eq:es1d}) can be solved as an initial value problem (IVP),
e.g., using a 4th-order Runge-Kutta scheme, which should produce the
exact linear Landau damping when the Case-Van Kampen mode and
numerical errors are ignored\cite{Xie2013}. This simulation approach
can be a simple and/or final benchmark for GPDF. Disagreements would
mean the GPDF has been treated incorrectly. However, the IVP
approach is not general, because of the numerical errors from
discrete of $v$ and $t$, especially when the phase velocity
$v_p=\omega/k$ or damping rate are large. A similar but more
complicated IVP approach is used by Valentini and
D'Agosta\cite{Valentini2007}. Particle (e.g., particle-in-cell)
simulation can also be used, which was also used by Godfrey {\it et
al.}\cite{Godfrey1975}. Particle method can also easily support
non-smooth distribution, but is limited by the noise resulting in
unfavorable errors. Thus, this technique is not accurate when
compared with the above continuum method.

Landau damping of Maxwellian distribution using this continuum IVP
approach is verified in a previous work\cite{Xie2013}.

\begin{figure}
  \includegraphics[width=8cm]{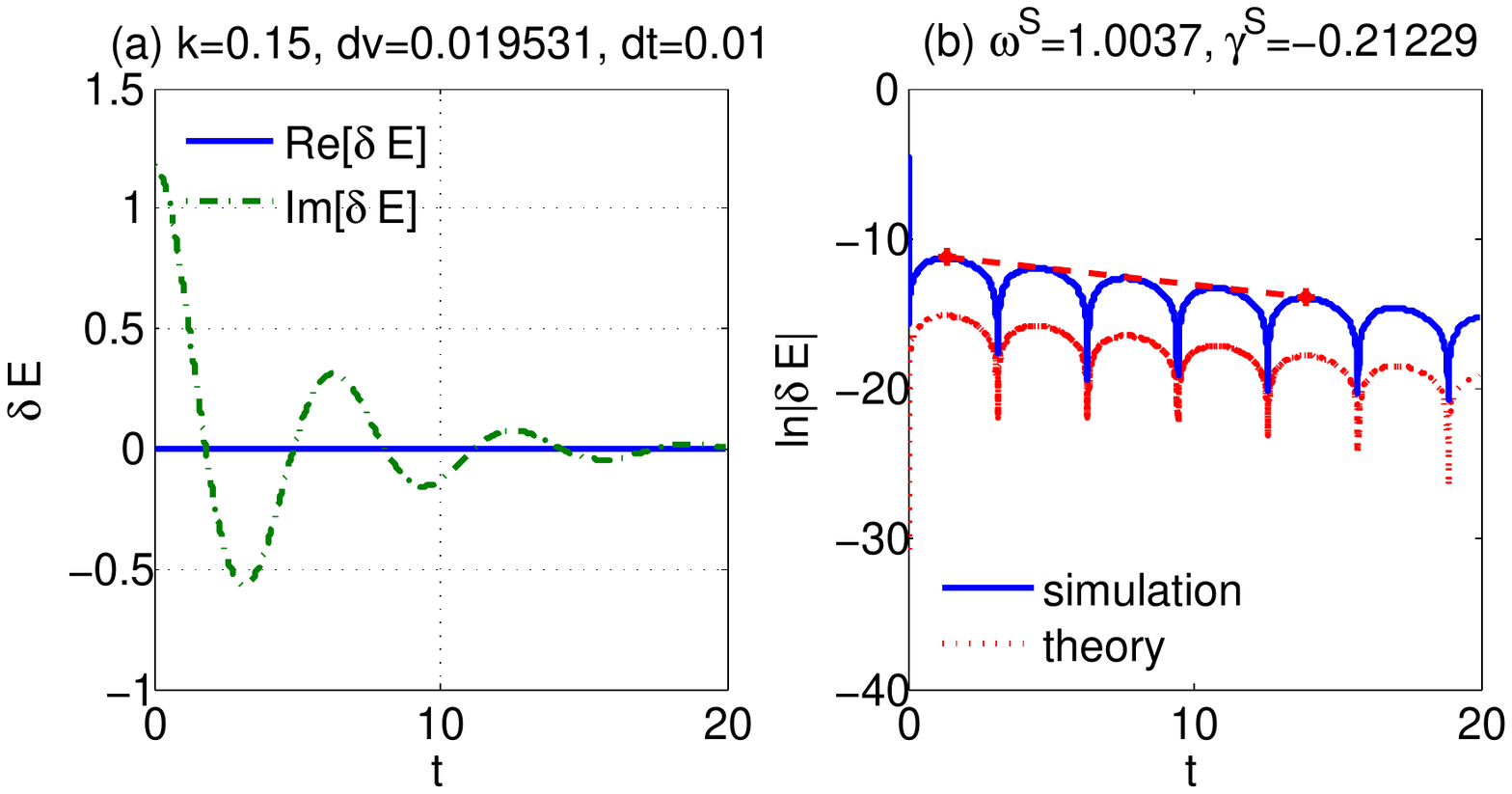}\\
  \caption{Benchmark GPDF using initial value simulation for
  Lorentzian distribution function. The result of root finding for this case is $\omega=1.0000-0.2121i$.}\label{fig:IVPLorentz}
\end{figure}

We check the Lorentzian distribution to show that our scheme for
GPDF with non-Maxwellian distributions is also correct.
Fig.\ref{fig:IVPLorentz} shows the comparison of Lorentzian
distribution Landau damping using IVP scheme and GPDF for $k=0.15$
and $v_t=\sqrt{2}$. We find that the simulation and
numerical/analytical solutions match very well for both real
frequency and damping rate, i.e., $\omega=1.00-0.212i$.

Another possible numerical approach for (\ref{eq:es1d}) is to treat
it as an eigenvalue problem. However, the (Landau) damping normal
mode is not eigenmode in this system, as discussed by numerous
authors (see e.g., \cite{Xie2013} and references in). Thus, this
approach does not work.

Notably, for GPDF, (\ref{eq:ZM}) would no longer be hold. Thus,
$Z_p$ should be used for root finding instead of $Z$ to
(\ref{eq:dres1d2}) or (\ref{eq:dres1d}).

\subsection{Effects of discontinuity point}
The $\delta$ function can be modeled as cold plasma, which provides
the dispersion relation $(\omega-kv_d)^2=\omega_p^2$. For
discontinuity point such as the cases we meet in incomplete
Maxwellian or flat top distributions, (\ref{eq:Zrect}) can be used
to solve the flat top case easily, which gives
$(\omega-kv_d)^2=\omega_p^2+k^2v_t^2$, where $v_t=(z_b-z_a)/2$ and
$v_d=(z_b+z_a)/2$. The dispersion relation solutions of $\delta$ and
flat top distributions are verified by PIC simulation (not shown
here). For instance, for flat top distribution $v_t=1.0$, $v_d=0$,
$k=1.0$, we obtain $\omega=1.414$, whereas PIC simulation yields
$\omega_r\simeq1.40$. For the continuum IVP simulation of
discontinuity point, we use the approximation $\partial f_0/\partial
v=[f_0(v_0^{+})-f_0(v_0^{-})]/\Delta v$, where $\Delta v$ is the
velocity space grid size. In a practical test, the simulation also
yields the same result, but is more accurate and has lower noise
than PIC simulation. Thus, we use continuum IVP to verify GPDF
results.

For incomplete Maxwellian distribution, we can compare the
dispersion relation solutions of $Z'_{\text{IM}}(\zeta)$ and
$Z'_{\text{IM0}}(\zeta)$ to investigate the effects of discontinuity
point.

\begin{figure}
  \includegraphics[width=8cm]{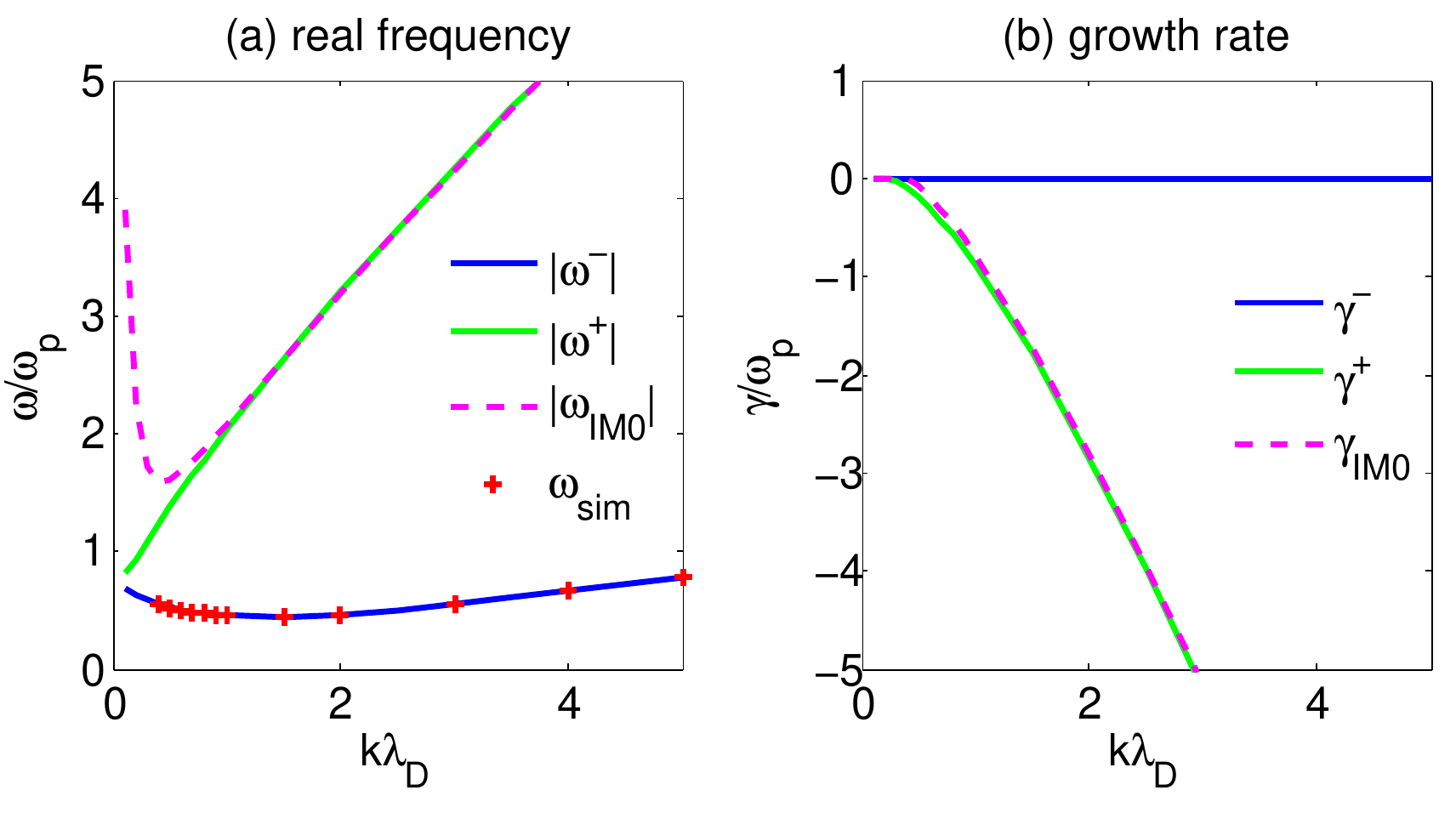}\\
  \caption{Dispersion relation solutions (root finding) of incomplete Maxwellian distribution
  with $\nu=-0.1v_t$, which agree with the simulation results very well.}\label{fig:DRIM}
\end{figure}

The results are shown in Fig.\ref{fig:DRIM}, with $\nu=-0.1v_t$ and
$v_t=\sqrt{2}$. For example, $k=1.0$, $Z'_{\text{IM}}$ yields
$\omega^{+}=$2.0409-i0.8801 and $\omega^{-}=$-0.4542+i3.5173E-5;
$Z'_{\text{IM0}}$ yields $\omega=$2.0843-i0.7871, whereas IVP
simulation yields $\omega\simeq$0.454+i0. Our solutions of
$\omega^{\pm}$ via GPDF are in favor of the solutions by
Baalrud\cite{Baalrud2013}. This benchmark provides further
verification of the one-solve-all scheme.

Fig.\ref{fig:DRIM} shows that the discontinuity point at $v=\nu$
changes the dispersion properties considerably, for instance, a new
nearly undamped branch can be found, which should be caused by the
lack of resonance particles at $v_p=\omega/k<\nu$. The differences
between the solutions of $Z'_{\text{IM}}(\zeta)$ and
$Z'_{\text{IM0}}(\zeta)$ also indicates that an incorrect treatment
of discontinuity point will yield inaccurate results.

\subsection{Results of distribution function effects}
We use GPDF to revisit the distribution function effects on Landau
damping.

\begin{figure}
  \includegraphics[width=8cm]{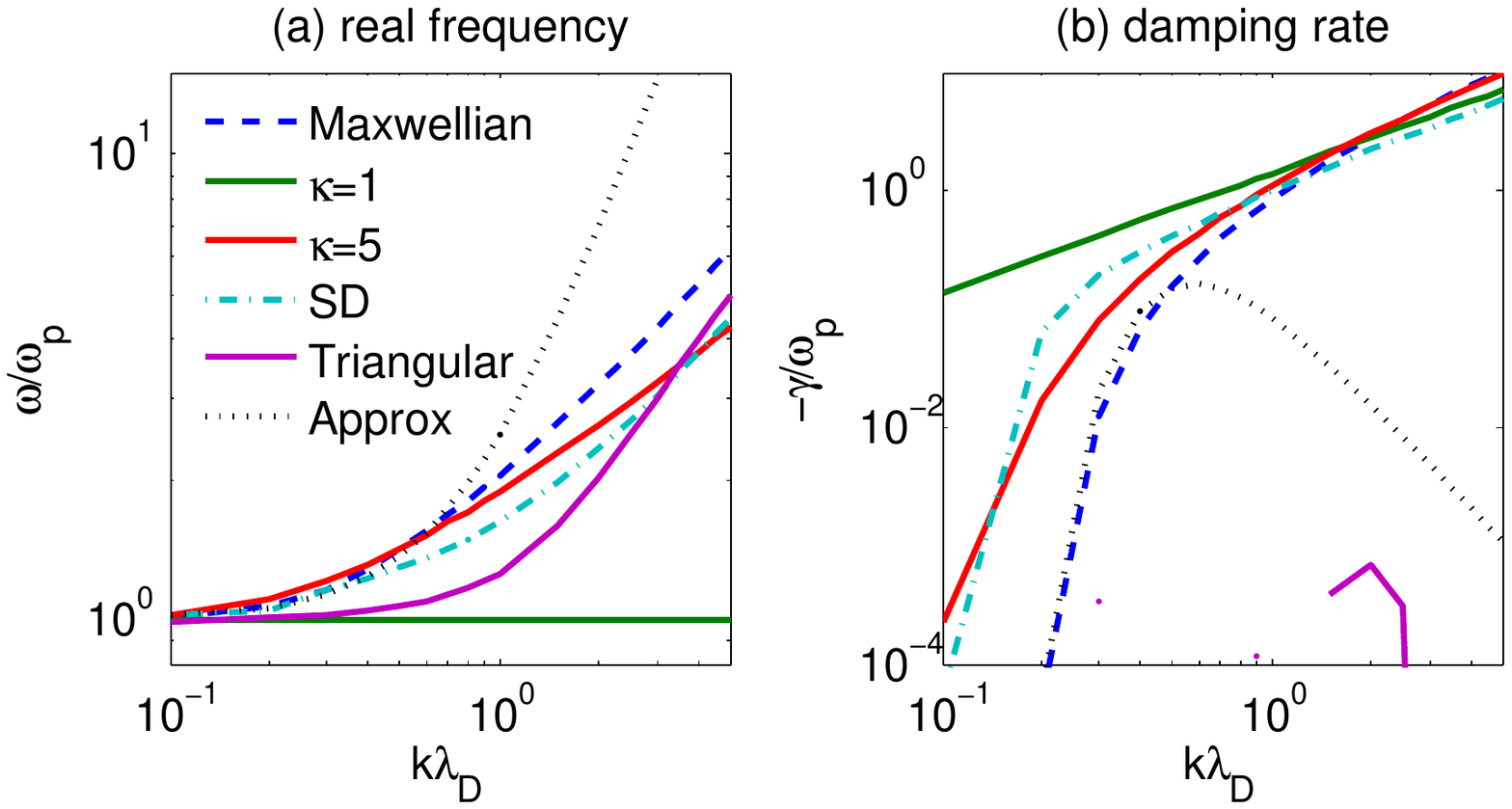}\\
  \caption{Distribution function effects on Landau damping with
  different initial distributions, where the approximate solution by (\ref{eq:wri}) for Maxwellian distribution is also shown.}\label{fig:LDRoot}
\end{figure}

For Maxwellian distribution and small $k$, approximate analytical
expressions for real frequency and growth rate
are\cite{Nicholson1983}
\begin{subequations}\label{eq:wri}
\begin{eqnarray}
  \omega_r^2=\omega_p^2(1+3k^2\lambda_D^2),\\
   \omega_i=\frac{\pi}{2}\frac{\omega_r^3}{k^2}\frac{\partial f_0}{\partial
  v}\Big|_{v=\omega_r/k}.
\end{eqnarray}
\end{subequations}

The effects of $\kappa$-distribution on Landau damping, particularly
for space plasma, are discussed in detail by Thorne and
Summers\cite{Thorne1991}. The incomplete Maxwellian distribution is
discussed by Baalrud\cite{Baalrud2013}.

\begin{table}
\caption{\label{tab:LD} Comparison of the Langmuir wave solutions
with different distribution functions and $k=1.0$.}
\begin{ruledtabular}
\begin{tabular}{ccccc}
  - & $\omega_r^G$ & $\omega_i^G$ & $\omega_r^S$ & $\omega_i^S$ \\\hline
  Maxwellian & 2.0459 & -0.8513 & 2.01 & -0.85 \\
  $\kappa=1$ & 1.0000 & -1.4142 & 1.00 & -1.40 \\
  $\kappa=5$ & 1.8786 & -1.0866 & 1.82 & -1.08 \\
  Slowing down & 1.6240 & -0.9975 & $\simeq$1.68 & $\simeq$-0.87 \\
  Triangular & 1.2577 & 4.2E-4 & $\simeq$1.26 & $\simeq$0.005 \\
\end{tabular}
\end{ruledtabular}
\end{table}

We choose Maxwellian, $\kappa$, and slowing down distributions with
$v_t=\sqrt{2}$ as well as triangular distributions with the same
parameters as in Fig.\ref{fig:ZTri} for further comparisons.
Fig.\ref{fig:LDRoot} shows the results of $\omega_r$ vs. $k$ and
$\gamma$ vs. $k$. For triangular distribution, a non-zero $\gamma$
around $\pm10^{-4}$ exists but is sensitive to initial guessing for
the root finding, which should be caused by the jump of $Z'$ around
$\Im(z)=0$ as shown in Fig.\ref{fig:ZTri}.

Table \ref{tab:LD} shows the quantitative value of the solutions,
where $\omega^G$ is solved from GPDF and $\omega^S$ is from IVP
simulation. For non-smooth input distributions, the simulation is
not robust and is sensitive to parameters and initial conditions.
Several results in table ($k=1.0$) are very rough (labeled with
`$\simeq$'). For instance, the error of the result in this table for
slowing down distribution is very large (approximately $10\%$),
whereas, for small $k$ and small damping rate, e.g., $k=0.5$, IVP
simulation is more robust and accurate, and we obtain
$\omega^S_{SD}=1.29-0.42i$ compared with
$\omega^G_{SD}=1.2876-0.4119i$.

\section{Summary and Discussion}\label{sec:summ}

The analytical properties and one-solve-all numerical scheme for
generalized plasma dispersion function, which provides a useful tool
for treating linear effects of almost arbitrary distribution
functions, are discussed. The exact distribution function effects on
Landau damping are revisited to demonstrate an application.

The one-solve-all scheme can also be used analytically as an
expansion method for GPDF, in addition to the usual Taylor expansion
scheme used.

Our method cannot be used directly for
relativistic\cite{Godfrey1975,Castejon2006} or other more
complicated dispersion functions because those dispersion functions
are usually not in HT form. However, similar orthogonal functions
expansion treatment may be used, as mentioned by
Robinson\cite{Robinson1990}.

\section{Acknowledgements}
Discussions with Y. R. Lin-Liu at the early stage of this project to
understand the treatment of usual PDF are appreciated. This work is
support by the NSF of China under Grants No.11235009, the ITER-CN
under Grant No. 2013GB104004 and Fundamental Research Fund for
Chinese Central Universities.

\end{CJK*}
\end{document}